\documentclass[smallextended]{svjour3}
\usepackage{graphicx, amsmath}
\usepackage[caption=false]{subfig} 

\newcommand{\ket}[1]{{\left| #1 \right\rangle}}

\journalname{Quantum Inf Process}

\begin{document}

\title{Diagrammatic Approach to Quantum Search}

\author{Thomas G.~Wong}

\authorrunning{T.~G.~Wong}

\institute{T.~G.~Wong \at
	   Faculty of Computing, University of Latvia, Rai\c{n}a bulv.~19, R\=\i ga, LV-1586, Latvia \\
	   \email{twong@lu.lv}
}

\date{Received: date / Accepted: date}

\maketitle

\begin{abstract}
	We introduce a simple diagrammatic approach for estimating how a randomly walking quantum particle searches on a graph in continuous-time, which involves sketching small weighted graphs with self-loops and considering degenerate perturbation theory's effects on them. Using this method, we give the first example of degenerate perturbation theory solving search on a graph whose evolution occurs in a subspace whose dimension grows with $N$.
\end{abstract}


\section{Introduction}

Degenerate perturbation theory is a ``textbook tool'' for quantum mechanics, famously used to derive the spectra of atoms in the presence of an external electric field (\textit{i.e.}, the Stark effect) \cite{Sakurai1993}. Recently, we showed that it can also be used to analyze quantum computing algorithms, specifically search on graphs by a randomly walking quantum particle evolving by Schr\"odinger's equation \cite{JMW2014}. Using it, we showed two intuitions to be false, that global symmetry and high connectivity are not necessary for fast quantum search \cite{JMW2014,MeyerWong2014}.

\begin{figure}
\begin{center}
	\includegraphics{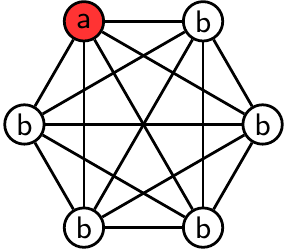}
	\caption{\label{fig:complete}Complete graph with $N = 6$ vertices.}
\end{center}
\end{figure}

For example, consider search on the complete graph with $N$ vertices, an example of which is shown in Fig.~\ref{fig:complete}. The vertices of the graph label computational basis states $\{ \ket{0}, \ket{1}, \dots, \ket{N-1} \}$ of an $N$-dimensional Hilbert space. Of these, we are looking for a particular ``marked'' vertex $\ket{a}$ using a randomly walking quantum particle, whose state $\ket{\psi(t)}$ begins in an equal superposition $\ket{s}$ of all the vertices:
\[ \ket{\psi(0)} = \ket{s} = \frac{1}{\sqrt{N}} \sum_{i = 0}^{N-1} \ket{i}. \]
It searches by evolving by Schr\"odinger's equation with Hamiltonian $H = -\gamma L - | a \rangle \langle a |$, where $\gamma$ is the jumping rate (\textit{i.e.}, amplitude per time), and $L = A - D$ is the graph Laplacian, which is composed of the adjacency matrix ($A_{ij} = 1$ if $i$ and $j$ are adjacent and $0$ otherwise) and the diagonal degree matrix ($D_{jj} = \deg(j)$ and $0$ otherwise) \cite{CG2004}. For a regular graph, $D$ is proportional to the identity matrix, so we can drop it by rezeroing the energy. Then the search Hamiltonian is
\begin{equation}
	\label{eq:H}
	H = -\gamma A - | a \rangle \langle a |.
\end{equation}

With this initial state and evolution, the non-marked vertices evolve identically by symmetry, as shown in Fig.~\ref{fig:complete}. So we can group them together:
\[ \ket{b} = \frac{1}{\sqrt{N-1}} \sum_{i \ne a} \ket{i}. \]
Then the system evolves in a two-dimensional subspace spanned by $\{\ket{a}, \ket{b}\}$, in which the Hamiltonian \eqref{eq:H} is
\begin{equation}
	\label{eq:H_complete}
	H = -\gamma \begin{pmatrix}
		\frac{1}{\gamma} & \sqrt{N-1} \\
		\sqrt{N-1} & N-2 \\
	\end{pmatrix}.
\end{equation}

\begin{figure}
\begin{center}
	\includegraphics{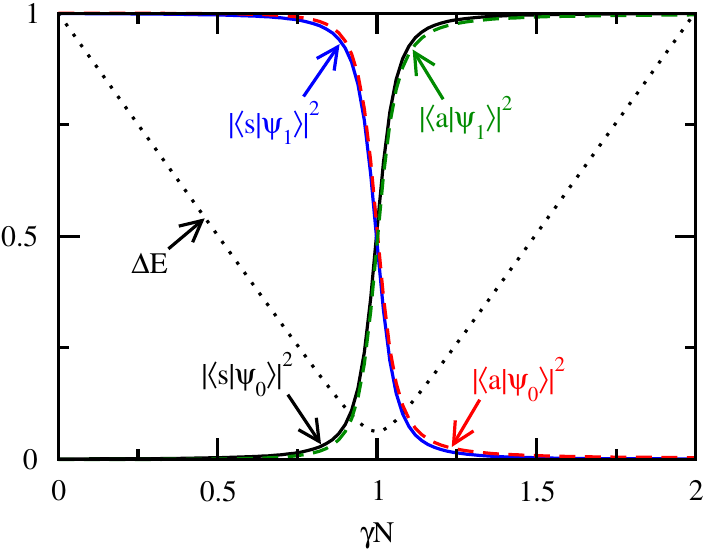}
	\caption{\label{fig:complete_overlap} Squared overlaps of the eigenstates of $H$ with $\ket{s}$ and $\ket{a}$ for the complete graph with $N = 1024$ vertices.}
\end{center}
\end{figure}

One traditionally estimates how the search algorithm evolves on a general graph by plotting the squared overlaps of the eigenstates of $H$ with $\ket{s}$, $\ket{a}$, and possibly other states \cite{MeyerWong2014,CG2004}. This is shown for the complete graph in Fig.~\ref{fig:complete_overlap}. From this, when $\gamma$ takes its critical value of $\gamma_c = 1/N$, the eigenstates of $H$ take the form $\ket{\psi_{0,1}} \propto \ket{s} \pm \ket{a}$, so the system evolves from $\ket{s}$ to $\ket{a}$ in time $\pi/\Delta E$.

To prove this and find the energy gap's scaling with $N$, we use degenerate perturbation theory \cite{JMW2014}. We begin by separating the Hamiltonian \eqref{eq:H_complete} into leading- and higher-order terms:
\[ H = \underbrace{-\gamma \begin{pmatrix}
	\frac{1}{\gamma} & 0 \\
	0 & N \\
\end{pmatrix}}_{H^{(0)}} + \underbrace{-\gamma \begin{pmatrix}
	0 & \sqrt{N} \\
	\sqrt{N} & 0 \\
\end{pmatrix}}_{H^{(1)}} + \,\dotsb. \]
In lowest order, the eigenstates of $H^{(0)}$ are $\ket{a}$ and $\ket{b}$ with corresponding eigenvalues $-1$ and $-\gamma N$. If the eigenvalues are nondegenerate, then the perturbation $H^{(1)}$ will not significantly change these eigenstates. Then since the initial superposition state $\ket{s}$ is approximately $\ket{b}$ for large $N$, the system will stay near its inital state, never having a large projection on $\ket{a}$. If the eigenstates are degenerate (\textit{i.e.}, when $\gamma$ takes its critical value of $1/N$), however, then the perturbation will cause the eigenstates of the perturbed system to be linear combinations of $\ket{a}$ and $\ket{b}$:
\[ \ket{\psi_{0,1}} = \alpha_a \ket{a} + \alpha_b \ket{b}, \]
where the coefficients $\alpha_{a,b}$ can be found by solving
\[ \begin{pmatrix} H_{aa} & H_{ab} \\ H_{ba} & H_{bb} \end{pmatrix} \begin{pmatrix} \alpha_a \\ \alpha_b \end{pmatrix} = E \begin{pmatrix} \alpha_a \\ \alpha_b \end{pmatrix}, \]
where $H_{ab} = \langle a | H^{(0)} + H^{(1)} | b \rangle$, \textit{etc}. Solving this yields $\ket{\psi_{0,1}} \propto \ket{b} \pm \ket{a}$ with corresponding eigenvalues $E_{0,1} = -1 \mp 1/\sqrt{N}$. Since $\ket{b} \approx \ket{s}$, the system evolves from $\ket{s}$ to $\ket{a}$ in time $\pi / \Delta E = \pi \sqrt{N} / 2$.

\begin{figure}
\begin{center}
\subfloat[]{
	\includegraphics{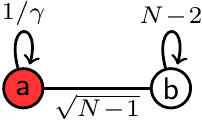}
	\label{fig:complete_diagram_H} 
} \quad \quad \quad \quad \quad
\subfloat[]{
	\includegraphics{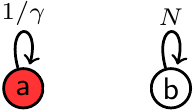}
	\label{fig:complete_diagram_H0} 
}
\caption{Apart from a factor of $-\gamma$, (a) the Hamiltonian for search on the complete graph represented as a weighted graph with self-loops, and (b) the leading-order terms.}
\end{center}
\end{figure}

This perturbative method can be interpreted to yield a simple diagrammatic approach to estimate how the search algorithm evolves, without needing to plot overlaps as in Fig.~\ref{fig:complete_overlap}. The search Hamiltonian \eqref{eq:H_complete} can be interpreted as the adjacency matrix of a weighted graph with two vertices and self-loops, as shown in Fig.~\ref{fig:complete_diagram_H}. The leading-order Hamiltonian $H^{(0)}$ is shown in Fig.~\ref{fig:complete_diagram_H0}, and it excludes the edge. Then the leading-order eigenstates are clearly $\ket{a}$ and $\ket{b}$. We choose $\gamma$ to make their eigenvalues degenerate so that, when the perturbation $H^{(1)}$ restores the missing edge, amplitude flows from $\ket{b}$ to $\ket{a}$. Since $\ket{s} \approx \ket{b}$, the system evolves from $\ket{s}$ to $\ket{a}$.

\section{Simplex of Complete Graphs}

\begin{figure}
\begin{center}
	\includegraphics{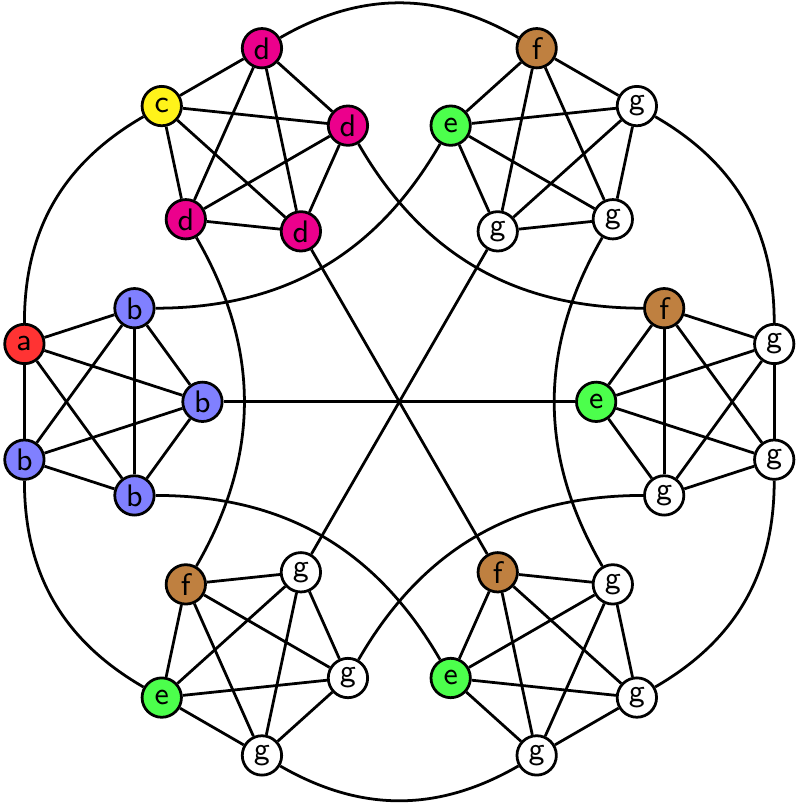}
	\caption{\label{fig:simplex} A 5-simplex with each vertex replaced with a complete graph of 5 vertices.}
\end{center}
\end{figure}

As a more complicated example of this diagrammatic approach, consider search on the $M$-simplex with each of its $M+1$ vertices replaced with a complete graph of $M$ vertices, an example of which is shown in Fig.~\ref{fig:simplex} \cite{MeyerWong2014}. As before, the $N = M(M+1)$ vertices of the graph label computational basis states, of which we are looking for a particular ``marked'' vertex $\ket{a}$ using a randomly walking quantum particle evolving by Schr\"odinger's equation with Hamiltonian \eqref{eq:H}.

In Fig.~\ref{fig:simplex}, identically evolving vertices are identically colored, and we see that the system evolves in a 7-dimensional subspace, independent of $M$ for $M > 2$. Grouping identically-evolving vertices, we get an orthonormal basis for this subspace:
\begin{align*}
	\ket{a} &= \ket{\text{red}} \\
	\ket{b} &= \frac{1}{\sqrt{M-1}} \sum_{i \in \text{blue}} \ket{i} \\
	\ket{c} &= \ket{\text{yellow}} \\
	\ket{d} &= \frac{1}{\sqrt{M-1}} \sum_{i \in \text{magenta}} \ket{i} \\
	\ket{e} &= \frac{1}{\sqrt{M-1}} \sum_{i \in \text{green}} \ket{i} \\
	\ket{f} &= \frac{1}{\sqrt{M-1}} \sum_{i \in \text{brown}} \ket{i} \\
	\ket{g} &= \frac{1}{\sqrt{(M-1)(M-2)}} \sum_{i \in \text{white}} \ket{i}.
\end{align*}
Then the Hamiltonian \eqref{eq:H} in this subspace is \cite{MeyerWong2014}
\[ \setlength{\arraycolsep}{2pt} H = -\gamma \! \left( \! \begin{matrix}
	\frac{1}{\gamma} & \sqrt{M-1} & 1 & 0 & 0 & 0 & 0 \\
	\sqrt{M-1} & M-2 & 0 & 0 & 1 & 0 & 0 \\
	1 & 0 & 0 & \sqrt{M-1} & 0 & 0 & 0 \\
	0 & 0 & \sqrt{M-1} & M-2 & 0 & 1 & 0 \\
	0 & 1 & 0 & 0 & 0 & 1 & \sqrt{M-2} \\
	0 & 0 & 0 & 1 & 1 & 0 & \sqrt{M-2} \\
	0 & 0 & 0 & 0 & \sqrt{M-2} & \sqrt{M-2} & M-2 \\
\end{matrix} \! \right) \! . \setlength{\arraycolsep}{3pt} \]

\begin{figure}
\begin{center}
\subfloat[]{
	\includegraphics{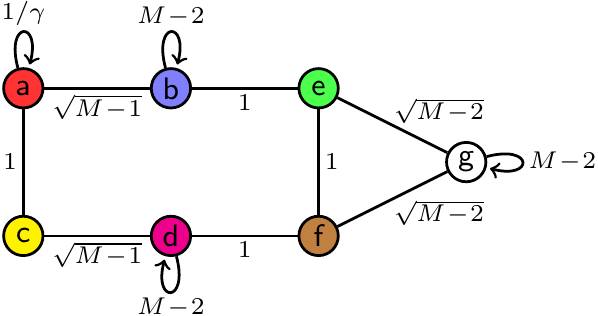}
	\label{fig:simplex_diagram_H}
}

\subfloat[]{
	\includegraphics{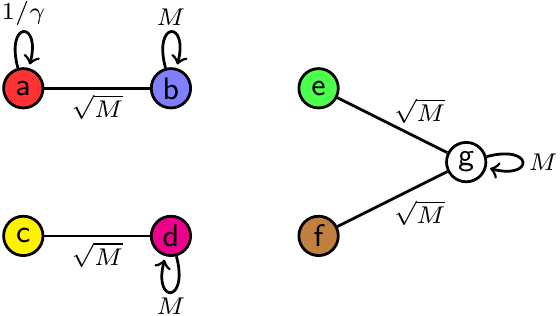}
	\label{fig:simplex_diagram_stage1}
}

\subfloat[]{
	\includegraphics{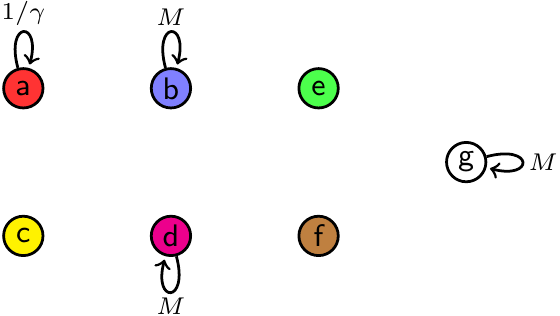}
	\label{fig:simplex_diagram_stage2}
}

\caption{\label{fig:simplex_diagram} Apart from a factor of $-\gamma$, (a) the Hamiltonian for search on the simplex of complete graphs represented as a weighted graph with self-loops, (b) the leading-order terms for the first stage of the algorithm, and (c) the leading-order terms for the second stage of the algorithm.}
\end{center}
\end{figure}

Using our diagrammatic approach, we can estimate the two-stage evolution of the algorithm without plotting overlaps as in Fig.~\ref{fig:complete_overlap} (and are available in \cite{MeyerWong2014}). In the 7-dimensional subspace, the Hamiltonian can be interpreted as the adjacency matrix of a weighted graph with seven vertices and four self-loops, as shown in Fig.~\ref{fig:simplex_diagram_H}. For the first stage of the algorithm, the leading-order Hamiltonian excludes the edges of weight 1, so we have Fig.~\ref{fig:simplex_diagram_stage1}. From this, we can visualize the seven eigenstates: two are superpositions of $\ket{a}$ and $\ket{b}$, two are superpositions of $\ket{c}$ and $\ket{d}$, and three are superpositions of $\ket{e}$, $\ket{f}$, and $\ket{g}$. We choose $\gamma$ so that the degenerate eigenstates are a superposition of $\ket{a}$ and $\ket{b}$ and a superposition of $\ket{e}$, $\ket{f}$, and $\ket{g}$. Then the perturbation restores the missing edges, and $\ket{g} \approx \ket{s}$ evolves to $\ket{b}$ since they are the most dominant pieces among $\ket{a}$, $\ket{b}$, $\ket{e}$, $\ket{f}$, and $\ket{g}$.

For the second stage of the algorithm, the leading-order Hamiltonian additionally excludes terms $\Theta(\sqrt{M})$, so we have Fig.~\ref{fig:simplex_diagram_stage2}. The seven eigenstates are simply $\ket{a}$, $\ket{b}$, \dots, $\ket{g}$. We choose $\gamma$ so that $\ket{a}$ is degenerate with $\ket{b}$, $\ket{d}$, and $\ket{g}$. Then the first-order perturbation restores the edges with weight $\Theta(\sqrt{M})$, giving us Fig.~\ref{fig:simplex_diagram_stage1}. Then probability at $\ket{b}$ will move to $\ket{a}$. So the overall evolution of both stages is to evolve from $\ket{s} \approx \ket{g}$ to $\ket{b}$ to $\ket{a}$, exactly as proved in \cite{MeyerWong2014}. Thus by sketching the small weighted graphs with self-loops in Fig.~\ref{fig:simplex_diagram}, we are able to estimate the evolution of the algorithm.


\section{Hypercube}

\begin{figure}
\begin{center}
	\includegraphics[width=1.5in]{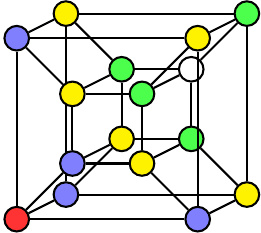}
	\caption{\label{fig:hypercube} 4-dimensional hypercube.}
\end{center}
\end{figure}

Degenerate perturbation theory has been used to solve quantum search problems on several different graphs, but they all evolved in constant-dimensional subspaces, namely 2D for the complete, 3D for strongly regular, 5D for joined complete, and 7D for the simplex of complete graphs \cite{JMW2014,MeyerWong2014}. Here we consider search on the $n$-dimensional hypercube, which has $N = 2^n$ vertices and evolves in an $(n+1)$-dimensional subspace. An example of this is shown in Fig.~\ref{fig:hypercube}. Although search on the hypercube was first solved in \cite{CG2004} using somewhat involved calculations from \cite{FGGS2000} and \cite{CDFGGL2002}, we solve it here much more simply using our diagrammatic approach as a guide. In doing so, we give the first example of degenerate perturbation theory solving a search problem where the evolution occurs in a subspace that grows with $N$.

\begin{figure*}
\begin{center}
\subfloat[]{
	\includegraphics{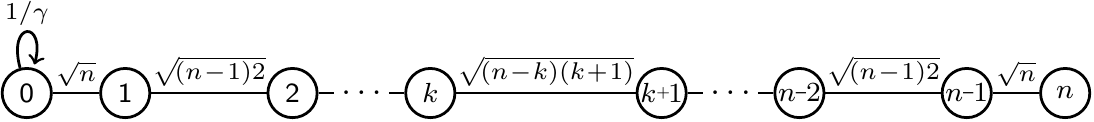}
	\label{fig:hypercube_diagram_H}
}

\subfloat[]{
	\includegraphics{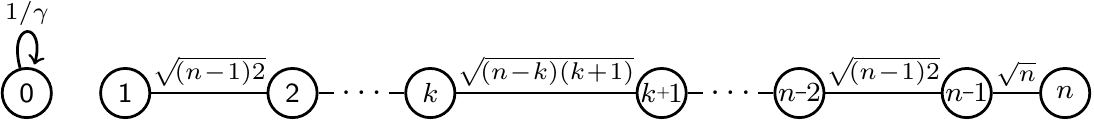}
	\label{fig:hypercube_diagram_H0}
}

\caption{Apart from a factor of $-\gamma$, (a) the Hamiltonian for search on the hypercube represented as a weighted graph with a self-loop, and (b) the leading-order terms.}
\end{center}
\end{figure*}

We begin by labeling each vertex with an $n$-bit string. Without loss of generality, we choose the marked vertex $\ket{a}$ to be the string of all 0's. Then vertices with the same number of 1's (\textit{i.e.}, with the same Hamming weight) evolve identically, and they can be grouped together:
\[ \ket{k} = \binom{n}{k}^{-1/2} \sum_{z_1 + \dots + z_n = k} \ket{z_1 \dots z_n}. \]
We use $\{\ket{k} : k = 0, 1, \dots, n\}$ as orthonormal basis vectors of the $(n+1)$-dimensional subspace. In this basis, the Hamiltonian \eqref{eq:H} is
\[ H \! = \! -\gamma \! \setlength{\arraycolsep}{1pt} \left( \!\! \begin{matrix}
	\frac{1}{\gamma} & \sqrt{n} \\
	\sqrt{n} & 0 & \ddots \\
	& \ddots & \ddots & \sqrt{(n\!-\!k)(k\!+\!1)} \\
	& & \sqrt{(n\!-\!k)(k\!+\!1)} & \ddots & \ddots \\
	& & & \ddots & 0 & \sqrt{n} \\
	& & & & \sqrt{n} & 0 \\
\end{matrix} \! \right) \!. \setlength{\arraycolsep}{3pt} \]
Using our diagrammatic approach, $H$ can be interpreted as a weighted graph with a self-loop, as shown in Fig.~\ref{fig:hypercube_diagram_H}. We choose the leading-order Hamiltonian to disconnect the marked vertex at the left end, yielding Fig.~\ref{fig:hypercube_diagram_H0}. Then $\ket{0}$ (\textit{i.e.}, $\ket{a}$) is an eigenvector of $H^{(0)}$ with eigenvalue $1/\gamma$ (ignoring the overall factor of $-\gamma$), and the remaining $n$ eigenvectors are linear combinations of $\ket{1}, \dots, \ket{n}$.

Note that without the self-loop, Fig.~\ref{fig:hypercube_diagram_H} represents the adjacency matrix of the $n$-dimensional hypercube, for which the equal superposition state $\ket{s}$ is an eigenstate with eigenvalue $n$ (so that $\ket{s}$ is an eigenvector of $L = A - D$ with eigenvalue $0$). Since for large $N$, this is approximately the vertices $\ket{1}, \dots, \ket{n}$ in Fig.~\ref{fig:hypercube_diagram_H0}, we expect the equal superposition over them
\[ \ket{r} = \frac{1}{\sqrt{N - 1}} \sum_{z_1 + \dots + z_n \ne 0} \ket{z_1 \dots z_n}, \]
which is approximately $\ket{s}$, to approximately be an eigenvector of $H^{(0)}$ with eigenvalue $n$. We see the accuracy of this approximation in Table \ref{table:hypercube_gamma}. For $n = 10$, our approximation that the reciprocal of the eigenvalue is $1/n$ already has four digits of accuracy.

Continuing with the diagrammatic approach, we make $\ket{r}$ degenerate with $\ket{a} = \ket{0}$, yielding $\gamma_c = 1/n$, so that the perturbation (\textit{i.e.}, restortation of the missing edge) causes the system to evolve from $\ket{s} \approx \ket{r}$ to $\ket{a}$. Table \ref{table:hypercube_gamma} compares our critical $\gamma$ with the more accurate value derived in \cite{CG2004} (corrected with an additional factor of $1/2$). For example, when $n = 40$, which is a ``database'' with about a trillion entries, the relative error is around 2.6\%. So while our method yields the correct asymptotic critical $\gamma$ with very little calculation, more careful analysis, such as that in \cite{CG2004}, may be needed for smaller graphs.

\begin{table}
\begin{center}
	\caption{\label{table:hypercube_gamma}Comparison of our critical $\gamma$ with \cite{CG2004} (corrected with an additional factor of $1/2$) for search on the $n$-dimensional hypercube.}
	\begin{tabular}{ccccc}
		\hline\noalign{\smallskip}
		$n$ & 1/(Actual Eig) & $1/n$ & \cite{CG2004} & Rel Error \\
		\noalign{\smallskip}\hline\noalign{\smallskip}
		10 & 0.100085 & 0.100000 & 0.114443 & 0.126201 \\
		20 & 0.050000 & 0.050000 & 0.052995 & 0.056517 \\
		30 & 0.033333 & 0.033333 & 0.034576 & 0.035934 \\
		40 & 0.025000 & 0.025000 & 0.025678 & 0.026398 \\
		50 & 0.020000 & 0.020000 & 0.020426 & 0.020873 \\
		60 & 0.016667 & 0.016667 & 0.016959 & 0.017264 \\
		70 & 0.014286 & 0.014286 & 0.014499 & 0.014719 \\
		80 & 0.012500 & 0.012500 & 0.012662 & 0.012829 \\
		90 & 0.011111 & 0.011111 & 0.011239 & 0.011370 \\
	       100 & 0.010000 & 0.010000 & 0.010103 & 0.010209 \\
		\noalign{\smallskip}\hline
	\end{tabular}
\end{center}
\end{table}

Doing the perturbative calculation, the eigenstates of the perturbed system are linear combinations of $\ket{a}$ and $\ket{r}$:
\[ \ket{\psi} = \alpha_a \ket{a} + \alpha_r \ket{r}, \]
where the coefficients can be found by solving
\[ \begin{pmatrix}
	H_{aa} & H_{ar} \\
	H_{ra} & H_{rr} \\
\end{pmatrix} \begin{pmatrix}
	\alpha_a \\
	\alpha_r \\
\end{pmatrix} = E \begin{pmatrix}
	\alpha_a \\
	\alpha_r \\
\end{pmatrix}, \]
where $H_{ar} = \langle a | H | r \rangle$, \textit{etc}. With $\gamma = 1/n$, this is for large $N$
\[ \begin{pmatrix}
	-1 & \frac{-1}{\sqrt{N-1}} \\
	\frac{-1}{\sqrt{N-1}} & -1 \\
\end{pmatrix} \begin{pmatrix}
	\alpha_a \\
	\alpha_r \\
\end{pmatrix} = E \begin{pmatrix}
	\alpha_a \\
	\alpha_r \\
\end{pmatrix}, \]
which has solutions $\ket{\psi_{0,1}} \propto \ket{r} \pm \ket{a}$ with corresponding eigenvalues $E_{0,1} = -1 \mp 1/(N-1)$. So the system evolves from $\ket{s} \approx \ket{r}$ to $\ket{a}$ in time $\pi/\Delta E = \Theta(\sqrt{N})$, which agrees with \cite{CG2004}.

From these examples, visualizing search by continuous-time quantum walk as small weighted graphs with self-loops provides a simple way to estimate the algorithm's evolution without needing to plot energy eigenstates. Using this diagrammatic approach to guide perturbative calculations, we see that degenerate perturbation theory's usefulness in analyzing quantum search on graphs is not restricted to problems that evolve in constant-dimensional subspaces, making it a more general tool for analyzing quantum search algorithms than one might be led to believe from its initial applications in \cite{JMW2014,MeyerWong2014}.


\begin{acknowledgements}
	Thanks to David Meyer for useful discussions. This work was partially supported by the European Union Seventh Framework Programme (FP7/2007-2013) under the QALGO (Grant Agreement No.~600700) project, and the ERC Advanced Grant MQC. 
\end{acknowledgements}


\bibliographystyle{my-spphys}
\bibliography{refs}

\end{document}